\documentclass[aps, prd, groupedaddress, preprint, tightenlines,  eqsecnum, nofootinbib, showpacs]{revtex4}
\usepackage{epsfig}
\usepackage{setspace}
\usepackage{graphicx}
\usepackage{leqno}
\usepackage{cancel}
\usepackage{amsmath}
\usepackage{amsfonts}
\usepackage{amssymb}
\usepackage{enumerate}

\usepackage{listings}
\usepackage{float, slashed, graphicx, amssymb, amsmath}
\usepackage[usenames, dvipsnames]{xcolor}
\usepackage{booktabs}

\usepackage[margin=2.2cm, a4paper, includefoot]{geometry}

\def\Li{\mathop{\hbox{\rm Li}}\nolimits}
\def\e{\epsilon}

\def\spa#1.#2{\left\langle#1\,#2\right\rangle}
\def\spb#1.#2{\left[#1\,#2\right]}
\def\la{\langle}
\def\ra{\rangle}

\newcommand{\tree}{\text{tree}}

\DeclareMathOperator{\tr}{\mathrm{tr}}

\def\st{I^{(1)}}

\newcommand\figref[1]{fig.~\ref{#1}}
\def\eps{\epsilon}

\def\be{\begin{equation}}
\def\ee{\end{equation}}

\begin{document}

\hfill\today

\title{Two-Loop Gravity amplitudes from four dimensional Unitarity} 

\author{David~C.~Dunbar, Guy R. Jehu and Warren~B.~Perkins}

\affiliation{
College of Science, \\
Swansea University, \\
Swansea, SA2 8PP, UK\\
\today
}

\begin{abstract}
We compute the polylogarithmic parts of the two-loop four and five-graviton amplitudes where the external helicities are 
positive and express these in a simple compact analytic form.  
We use these to extract the $\ln(\mu^2)$ terms from both the four and five-point amplitudes and show that these match the same
 $R^3$ counterterm.  
\end{abstract}


\maketitle

\section{Introduction}

Computing the scattering amplitudes of a quantum theory using its singular and analytic structure has a long history with many notable sucesses~\cite{Eden}. 
Recently the five-point all-plus helicity amplitude has been computed in QCD: first the integrands were determined 
using the method of maximal cuts~\cite{Badger:2013gxa} and $D$-dimensional unitarity and then the integrals were evaluated yielding a compact analytic form~\cite{Gehrmann:2015bfy}.  In $D$-dimensional unitarity the cuts are computed in $D=4-2\eps$ dimensions where, typically, the
components of the cuts are considerably more complicated than in four dimensions.
In~\cite{Dunbar:2016aux} it was demonstrated that four-dimensional unitarity techniques~\cite{Bern:1994zx,Bern:1994cg} blended with a knowledge of the singular structure of the amplitude could reproduce this form in a 
straightforward way.  The four-dimensional approach was also used to calculate the 
six-point amplitude~\cite{Dunbar:2016cxp,Dunbar:2016gjb} which was subsequently verified~\cite{Badger:2016ozq}.

Here we apply these techniques to gravity amplitudes with a particular emphasis on their Ultra-Violet (UV) behaviour. 
Understanding the UV structure of quantum gravity necessitates studying the two-loop amplitude. 't Hooft and Veltman~\cite{OneLoopFinite} demonstrated the one-loop finiteness of on-shell amplitudes in quantum gravity by showing the available counterterms in four dimensions made no contribution to perturbative amplitudes.  This cancellation does not persist to two loop and Goroff and Sagnotti~\cite{Goroff:1985th,vandeVen:1991gw} in a landmark calculation
were able to compute a UV infinity of the form
\begin{equation}
\sim { 209\over 80 \epsilon} \times  A^R
\end{equation}
where $A^R$ is the functional form generated by an $R^3$ counterterm.  
A feature of the computation was the appearance of sub-divergences which were cancelled diagram by diagram. 
In~\cite{Bern:2015xsa} this computation was revisited using evanescent operators which arise at one-loop to 
remove the sub-divergences in the four-point two-loop amplitude. 
For gravity coupled to various matter multiplets they found UV terms 
\begin{equation}
\left(  { a_1\over  \epsilon} +  a_2 \ln(\mu^2) \right) \times  A^R
\end{equation}
and noted that the coefficient of $\ln(\mu^2)$, $a_2$,  was more robust and simpler than $a_1$. Specifically, when coupled to $n_3$  (non-propagating) 
three form fields $a_1$ obtained a contribution proportional to $n_3$ but $a_2$ did not.  Additionally when coupled to 
scalars and vectors  $a_2$ was simply proportional to the the difference between the number of bosonic and fermionic degrees of freedom. 
As was argued in ref~\cite{Bern:2015xsa}, the coefficient $a_2$ has physical content since after renormalisation the amplitude depends upon $a_2$ but not $a_1$.  

In this article we explore and compute, up to rational terms, the four and five-point two-loop amplitude in quantum gravity where all the external gravitons  
have positive helicity. We use the techniques which have proven successful for the gluon amplitude: the amplitude is organised using a knowledge of its 
singularity structure, then four-dimensional unitarity is used to determine the logarithmic and dilogarithmic parts. 
Since $\ln(\mu^2)$ only appears in the combination $\ln(K^2/\mu^2)$  we can extract the coefficient of 
$\ln(\mu^2)$ using unitarity. We obtain the same coefficient of $\ln(\mu^2)$ for the four-point amplitude as in ref.~\cite{Bern:2015xsa}.
We also obtain the coefficient of $\ln(\mu^2)$ for the five-point amplitude and show that this  matches to the same  $R^3$ counterterm.

\section{Structure of the Amplitude}
\def\rg{{\rm r}_{\Gamma}}

As a convention we remove the coupling constant factors from the full $n$-point $L$-loop amplitude, ${\cal M}_n^{(L)}$ using
\begin{equation}
{\cal M}_n^{(L)}(1,\cdots   ,n)  = { i(\kappa/2)^{n-2+2L} (\rg)^L \over (4\pi)^{L(2-\epsilon)}  } {M}_n^{(L)}(1,\cdots   ,n)
\label{eq:fullampdefinition}
\end{equation}
where $\rg=\Gamma^2(1-\eps)\Gamma(1+\epsilon)/\Gamma(1-2\epsilon)$.

We then organise the  amplitude according to its singularity structure.  
The amplitude has both Infra-Red (IR) and UV singularities in the dimensional regulation parameter $\epsilon$.  
The all-plus amplitudes, which are finite at one-loop, can be divided as:
\begin{align}
M^{(2)}_{n} =& M^{(1)}_{n}\st_n  +\;G^{(2)}_n  + \;F^{(2)}_{n}  +\;R^{(2)}_{n} + {\mathcal O}(\eps)\,,
\label{eq:definitionremainder}
\end{align}
where the first term  contains the IR singularities of the amplitude. The function $\st_n$ is~\cite{Weinberg:1965nx,Dunbar:1995ed,Naculich:2011ry,Akhoury:2011kq}
\begin{align}\label{definitionremainderI}
\st_{n} =& \left[ - \sum_{i < j}^{n} \frac{1}{\epsilon^2}  s_{ij}\left( \frac{\mu^2}{-s_{ij}}\right)^{\epsilon}  \right]  
\end{align}
and $M^{(1)}_{n}$ is the all-$\epsilon$ form of the one-loop amplitude (with $s_{ij}=(k_i+k_j)^2$).
The leading singularity of $\st_n$ is only $\eps^{-1}$ since 
\begin{equation}
{1 \over \epsilon^2} \left( \sum_{i<j}^n s_{ij} \right)={1 \over \epsilon^2} \times 0
\end{equation}
by momentum conservation and the leading singularity is then
\begin{equation}
{1 \over \epsilon} \times \left( \sum_{i<j}^n s_{ij}\ln(-s_{ij}/\mu^2 ) \right) \times M_n^{(1)} \,.
\end{equation}
The amplitude also has finite logarithmic terms. In our four dimensional formulation these arise in two-particle cuts and have 
the form of one-loop bubble integral functions,
\begin{equation}
\;G^{(2)}_n =\sum_{i < j}^{n} c_{ij} \frac{1}{\epsilon} \left(\frac{\mu^2}{-s_{ij}}\right)^{\epsilon}\,. 
\end{equation}
where $c_{ij}$ are rational functions of $\lambda_k$ and $\bar\lambda_k$.\footnote{As usual,  a null momentum is represented as a
pair of two component spinors $p^\mu =\sigma^\mu_{\alpha\dot\alpha}
\lambda^{\alpha}\bar\lambda^{\dot\alpha}$. For real momenta
$\lambda=\pm\bar\lambda^*$ but for complex momenta $\lambda$ and
$\bar\lambda$ are independent~\cite{Witten:2003nn}. We are using a spinor helicity formalism with the usual
spinor products  $\spa{a}.{b}=\epsilon_{\alpha\beta}
\lambda_a^\alpha \lambda_b^{\beta}$  and 
 $\spb{a}.{b}=-\epsilon_{\dot\alpha\dot\beta} \bar\lambda_a^{\dot\alpha} \bar\lambda_b^{\dot\beta}$. }

The all-plus two-loop amplitude in QCD does not contain this term~\cite{Catani:1998bh}.
This give rise to $\epsilon^{-1}$ and $\ln(\mu^2)$ terms in the combination
\begin{equation}
\left( \sum_{i<j} c_{ij} \right) \times \left( {1\over \eps}+\ln(\mu^2) \right).
\end{equation}
There may be other sources of  $\eps^{-1}$ terms not directly determined by unitarity.  
The function $F^{(2)}_{n}$ contains the remaining polylogarithms of the amplitude and $R^{(2)}_{n}$ contains the
remaining rational terms. 
In dimensional regularisation
the internal momenta lie in $D=4-2\eps$ and it is really $D$-dimensional unitarity which should be used to reconstruct the amplitude.
Consequently, four dimensional unitarity is not sensitive to the rational terms, however it does give considerable simplification. 
The rational terms may be computed by complementary methods such as recursion.   
For the case of QCD, the rational terms for $n=5,6$ were computed by recursion starting from the four point amplitude.  We will not compute $R^{(2)}_5$ here: it not being necessary for our analysis and $R^{(2)}_4$ not being available.

The all-plus two-loop amplitude is a particularly simple amplitude. The all-plus helicity tree amplitude vanishes,
\begin{equation}
M^{(0)}_n(1^+,2^+,\cdots ,n^+) = 0
\end{equation}
This can be seen as a consequence of supersymmetric Ward identities~\cite{SWI}.  These imply that this amplitude vanishes to all orders in perturbation theory in supersymmetric theories. Since the $n$-graviton amplitudes for pure gravity coincide with those for supersymmetric theories at tree level then the gravity tree amplitude also vanishes.

\begin{figure}[h]
    \includegraphics{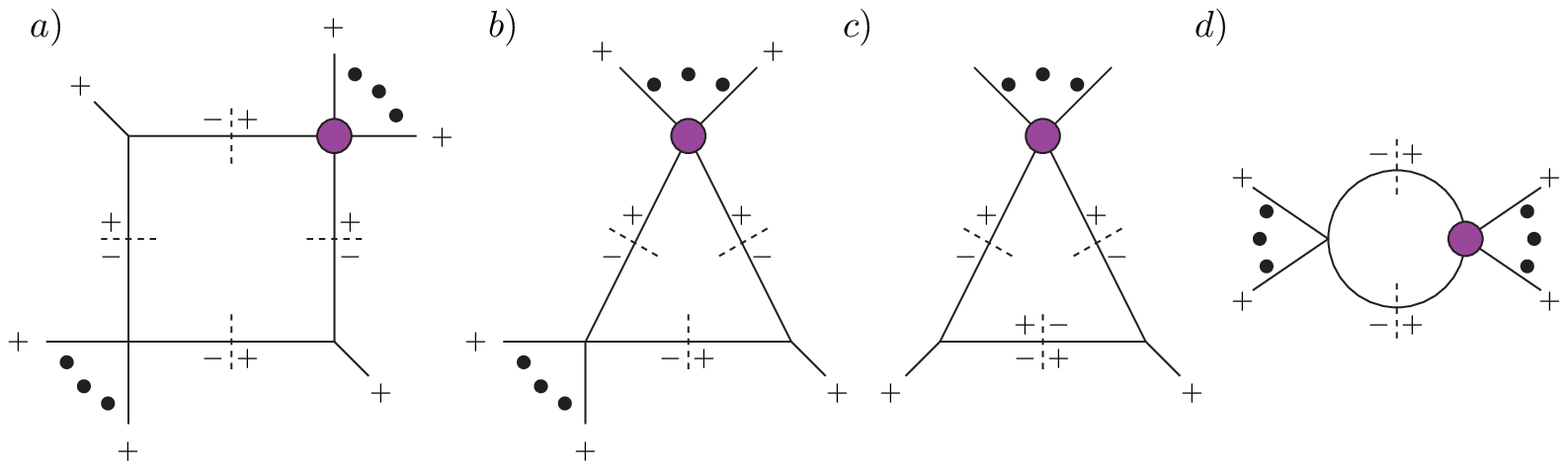}
   \caption{Four dimensional cuts of the two-loop all-plus amplitude involving an all-plus one-loop vertex 
    (indicated by $\bullet\;$ )} 
    \label{fig:oneloopstyle}
\end{figure}

The one-loop four-point amplitude for pure gravity is~\cite{Dunbar:1994bn}\footnote{We use for four point kinematics 
$s \equiv s_{12}$, $t\equiv s_{14}$ and $u\equiv s_{13}$.}
\begin{equation}
M_4^{(1)}(1^+,2^+,3^+,4^+)=  -
\left( {s t \over \spa1.2\spa2.3\spa3.4\spa4.1 } \right)^2 { (s^2+t^2+u^2) \over 120 }
+O(\epsilon)
\end{equation}
and the $n$-point amplitude can be expressed as~\cite{Bern:1998sv}
\begin{align}
M_n^{(1)}(1^+,2^+,3^+,\cdots ,n^+)= \frac{(-1)^n}{960} \sum_{a,b,M,N}  h(a,M,b)h(b,N,a) \tr^3(aMbN)
+O(\epsilon)
\end{align}
where $h(a,M,b)$ are the ``half-soft'' functions of ref.~\cite{Bern:1998sv}.  The summation is over pairs of legs $(a,b)$ and 
partitions $(M,N)$ of the remaining legs where both the sets $M$ and $N$ have at least one element. 
The half-soft functions we need for the five-point amplitude are
\begin{equation}
h(a,\{c\},b)={ 1\over \spa{a}.c^2 \spa{c}.b^2} \; , 
\;\;\;
h(a,\{c,d\},b)={ 1\over \spa{a}.c\spa{a}.d} {\spb{c}.{d} \over \spa{c}.{d} }{1\over  \spa{c}.b\spa{d}.b} 
\; . 
\end{equation}
When coupled minimally to additional bosons and fermions these  amplitudes are multiplied by a factor 
$(N_B-N_F)/2$ where $N_{B/F}$ is the total number of bosonic/fermionic degrees of freedom.  
A key feature of the one-loop amplitudes is that they are, to order $\epsilon^0$,  rational functions and as such have no 
cuts in four dimensions. Thus if computing amplitudes using cuts in four dimensions they are 
indivisible and can be treated as a vertex. The only four dimensional cuts of the $n$-point all-plus amplitude are shown in fig.~\ref{fig:oneloopstyle}.  For this helicity amplitude the only tree amplitudes necessary to compute the cuts are the three point amplitudes and the
Maximally-Helicity-Violating (MHV) tree amplitudes~\cite{Berends:1988zp}.


The four-dimensional calculation gives the coefficient of $I_n^{(1)}$ to be $M_n^{(1)}$ to leading order in $\epsilon$.  As in the QCD case, 
we  promote this to the all-$\epsilon$ form. 
The all-plus one-loop amplitudes in eq.~(\ref{eq:definitionremainder}) for four and five-points are known to all orders in $\epsilon$~\cite{Bern:1996ja,Bern:1998sv}
\def\mmu{\mu_0}
\begin{align}
M_4^{(1)}(1^+,2^+,3^+,4^+)&=2 { \spb1.2^2\spb3.4^2\over \spa1.2^2\spa3.4^2}
\left( I_4^{1234}[\mmu^8]+I_4^{1243}[\mmu^8]+I_4^{1423}[\mmu^8] \right)
\notag
\\
M_5^{(1)}(1^+,2^+,3^+,4^+,5^+)&=\beta_{123(45)} I_4^{123(45)}[\mmu^8]
-2{ \spb1.2\spb2.3\spb3.4\spb4.5\spb5.1\over\spa1.2\spa2.3\spa3.4\spa4.5\spa5.1} I_5^{12345}[\mmu^{10}]+{\rm Perms}
\end{align}
where
\begin{equation}
\beta_{123(45)}=-\spb1.2^2\spb2.3^2 h(1,\{4,5\},3)
\end{equation}
and $I_n[\mmu^{2m}]$ are the $n$-point integral functions with $\mmu^{2m}$ inserted where $\mmu$ are the $-2\eps$ coordinates in dimensional regularisation,
$\int d^Dx=\int d^4x d^{-2\eps} \mmu$.  The superscripts denote the ordering and clustering of the external legs. 
These are related to scalar integrals in higher dimensions~\cite{Bern:1995db,Bern:1996ja},
\begin{equation}
I_m[\mmu^{2r}]=-\epsilon(1-\epsilon)\cdots (r-1-\epsilon)(4\pi)^r I_m^{D=4+2r-2\epsilon} 
\; . 
\end{equation}

\section{The Four-Point All-Plus Helicity Amplitude}

The four point all-plus helicity amplitude has some significant simplifications. Specifically, the quadruple cuts vanish\footnote{In computing the quadruple cuts 
for a four-point amplitude the only non-vanishing product of cut amplitudes has alternating $MHV$ and $\overline{MHV}$ three-point vertices at the corners. This precludes any box functions for the four-point all-plus amplitude.}
and there are only one-mass triangle and bubble contributions.  In fact this amplitude is sufficiently simple that using the one loop amplitude as a vertex both the triangle and bubble  functions can be obtained simply from the 
two-particle cuts~\cite{Dunbar:1994bn} with the result,
\begin{align}
M^{(2)}_4(1^+,&2^+,3^+,4^+)
=M^{(1)}_4(1^+,2^+,3^+,4^+)\times \biggl( { 2(-s)^{1-\epsilon}+2(-t)^{1-\epsilon}+2(-u)^{1-\epsilon} \over \epsilon^2 }
\notag \\
&-{2s(3u^2+3t^2-2s^2) \over (s^2+t^2+u^2)}{ (-s/\mu^2)^{-\eps} \over \eps }  
-{2t(3s^2+3u^2-2t^2) \over (s^2+t^2+u^2)}{ (-t/\mu^2)^{-\eps} \over \eps }
\notag \\
&-{2u(3t^2+3s^2-2u^2) \over (s^2+t^2+u^2)}{ (-u/\mu^2)^{-\eps} \over \eps }
+\hbox{\rm rational terms} \biggr) \,.
\end{align}
This expression contains  $\eps^{-1}$ and the $\ln(\mu^2)$ terms in the combination, 
\begin{align}
& M^{(1)}_4(1^+,2^+,3^+,4^+)\times  { 30 stu \over (s^2+t^2+u^2) } \times \left( \frac{1}{\epsilon}+\ln(\mu^2) \right)
\notag
\\
=&-
{ 1 \over 4 }
\left( {s t \over \spa1.2\spa2.3\spa3.4\spa4.1 } \right)^2 \times  { stu } \times \left( \frac{1}{\epsilon}+\ln(\mu^2) \right)\,.
\label{eq:fourpoint}
\end{align}

If gravity is coupled to matter with $N_B-2$ additional bosonic degrees of freedom and $N_F$ fermionic degrees of freedom, the one-loop all-plus amplitude is  multiplied by a factor of $(N_B-N_F)/2$ and 
the calculation follows through as above. The all-plus amplitude in this theory is thus the expression above with the replacement
\begin{equation}
-\frac{1}{4} \longrightarrow -{ N_B-N_F \over 8} 
\end{equation}
which matches the result of \cite{Bern:2015xsa}. 
Consequently, we note that four dimensional unitarity gives the correct ${\it renormalised}$ amplitude up to rational terms although, as was known in ref~\cite{Dunbar:1994bn},
the coefficient of $\epsilon^{-1}$ does not match the field theory calculation.

\section{The Five Point All-Plus Amplitude}

\begin{figure}[h]
    \includegraphics{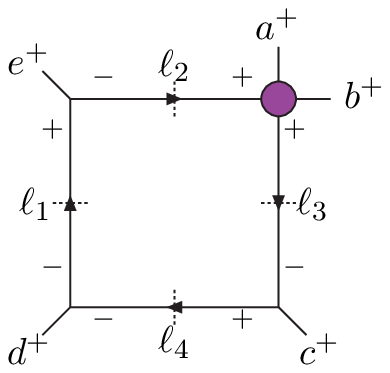}
    \caption{The labelling and internal helicities of the quadruple cut.}
    \label{fig:oneloopbox} 
\end{figure}

This amplitude contains functions, particularly dilogarithms, that are not present in the four-point amplitude.  These are contained in 
the box contributions shown in fig.~(\ref{fig:oneloopbox}). 
The box contribution is readily evaluated using a quadruple cut~\cite{BrittoUnitarity}. 

With the labelling of \figref{fig:oneloopbox} the cut momenta are
\begin{align}
\ell_1={\spa{c}.d\over\spa{e}.c} \bar\lambda_d \lambda_e
\,\, &,\,\,
\ell_2={\la c|P_{de}|\over\spa{e}.c}  \lambda_e
\,\, ,\,\, 
 \ell_3={\la e|P_{cd}|\over\spa{e}.c}  \lambda_c
\,\, ,\,\, 
 \ell_4={\spa{e}.d\over\spa{e}.c} \bar\lambda_d \lambda_c
 \,\, ,\,\, 
\end{align}
giving the coefficient of the box function
\begin{align}
{\cal C}_{ \{a,b\}, c,d,e }=& \frac{1}{2}
M_4^{(1)} (a^+,b^+,\ell_3^+,-\ell_2^+) \times  M_3^{(0)}( -l_3^-, c^+ ,l_4^+ ) \times  M_3^{(0)}(-l_4^-, d^+, l_1^-)  \times 
M_3^{(0)}(-l_1^+, e^+, l_2^-)
\notag
\\
=&{1\over 240}  \left( { \spb{a}.{b}^6 \spb{c}.d^2  \spb{e}.d^2 \over \spa{e}.c^4 } \right)  
\left( {\spa{a}.{b}^2   \spa{e}.c^2 }  + {  \spa{a}.e^2\spa{b}.c^2   } + { \spa{a}.c^2\spa{b}.e^2   }  \right) 
\end{align}

This is the coefficient of the integral function $I_4^{\rm 1m}(s_{cd},s_{de},s_{ab})$ where~\cite{Bern:1993kr}
\def\rg{r_{\Gamma}}
\def\e{\epsilon}
\def\Li{{\rm Li}}
\def\mum2{(\mu^{2\epsilon})}
\begin{align}
I^{\rm 1m}_4(S,T,M^2) =& -{2  \over S T }
\Biggl[-{\mum2\over\e^2} \Bigl[ (-S)^{-\e} +
(-T)^{-\e} - (-M^2)^{-\e} \Bigr] \cr
\notag \\ 
  + &\Li_2\left(1-{ M^2 \over S }\right)
   + \ \Li_2\left(1-{M^2 \over T}\right)
   +{1\over 2} \ln^2\left({ S  \over T}\right)
+\ {\pi^2\over6}
\Biggr] +{\cal O}(\epsilon)
\end{align}
and overall factors have been  removed according to the normalisation of eq.~(\ref{eq:fullampdefinition}).

This integral function splits into singular terms plus a remainder $I_4^{\rm 1m}=I_4^{\rm 1m:IR}+I_4^{\rm 1m:F}$
where
\begin{equation}
I_4^{\rm 1m:IR}(S,T,M^2) \equiv  -{2\over S T }
\Biggl[-{\mum2 \over\e^2} \Bigl[ (-S)^{-\e} +
(-T)^{-\e} - (-M^2)^{-\e} \Bigr] \Biggr]
\; . 
\end{equation}

The one-mass integral function is\begin{align}
I_{3}^{1\rm m}(K^2) = { \mum2 \over\e^2} (-K^2)^{-1-\e}.
\end{align}
and the two-mass triangle function is,
\begin{align}
I_{3}^{2 \rm m}\bigl(K_1^2,K_2^2\bigr)= {\mum2 \over\e^2}
{(-K_1^2)^{-\e}-(-K_2^2)^{-\e} \over  (-K_1^2)-(-K_2^2) }\ .
\end{align} 
The boxes, one-mass and two-mass triangles all have IR infinite terms of the form
\begin{equation}
{\mum2 \over\e^2}(-K^2)^{-\e}
\end{equation}
The coefficients of the triangle contributions can be evaluated using triple cuts~\cite{Bidder:2005ri,Darren,BjerrumBohr:2007vu,Mastrolia:2006ki}   
and a canonical basis~\cite{Dunbar:2009ax}.  
Summation over the box and triangle 
contributions gives an
overall coefficient of $M^{(1),\epsilon^0}_5(a^+,b^+,c^+,d^+,e^+)$, i.e
\begin{align}
&
\left(   \sum  {\cal C}_{\{a,b\},c,d,e} I_4^{1\rm m}
+\sum  {\cal C}_{\{a,b,c\},d,e} I_{3}^{1\rm m}
+\sum  {\cal C}_{\{a,b\},c,\{d,e\} }  I_{3}^{2 \rm m}  \right)_{IR}
\notag \\ = &
M^{(1),\epsilon^0}_5(a^+,b^+,c^+,d^+,e^+)
\times 
\sum_{i < j}^{n}  -\frac{1}{\epsilon^2}  s_{ij} \left(\frac{\mu^2}{-s_{ij}}\right)^{\epsilon} 
\label{eq:sumofIR}
\end{align}
where $M^{(1),\epsilon^0}_5(a^+,b^+,c^+,d^+,e^+)$ is the order $\epsilon^0$ truncation of the one-loop amplitude.
A key step is to promote the coefficient of these terms to the all-$\epsilon$ form of the one-loop amplitude which 
then gives the correct singular structure of the amplitude. 
We have confirmed relationship~(\ref{eq:sumofIR}) for the $n$-point amplitude by computing the triple and quadruple  cuts at specific kinematic points up to $n=10$. 

Consequently, we can obtain the compact explicit analytic form for the dilogarithmic remainder part of the amplitude
\begin{align}
F_5^{(2)}=
{1\over 240}
\sum_{P_{30}}
\left( { \spb{a}.{b}^6 \spb{c}.d^2  \spb{e}.d^2 \over \spa{e}.c^4 } \right)  
\biggl( {\spa{a}.{b}^2   \spa{e}.c^2 }  + {  \spa{a}.e^2\spa{b}.c^2   } + { \spa{a}.c^2\spa{b}.e^2   }  \biggr) 
\notag
\\
\times  \Biggl( 
 -{2 \over s_{cd} s_{de}  } \Biggr)
\Biggl[ 
\Li_2\left(1-{ s_{ab} \over s_{cd} }\right)
   +  \Li_2\left(1-{s_{ab}  \over s_{de} }\right)
 +{1\over 2} \ln^2\left(     {  s_{cd}  \over s_{de} }\right)
+ {\pi^2\over6}
\Biggr]  \; ,
\label{eq:cc}
\end{align}
where the permutation sum is over the 30 independent permutations of the legs $(\{a,b\},c,d,e)$ after factoring out for the
symmetries $(\{a,b\},c,d,e)\equiv (\{b,a\},c,d,e)\equiv (\{a,b\},e,d,c)$.

Note that the coefficients in the $F_5^{(2)}$ term contain $\spa{e}.c^{-4}$ singularities. On  this singularity the integral function vanishes and $F_5^{(2)}$ has
$\spa{e}.c^{-3}$ singularities.  These are spurious and not present in the full amplitude. They cancel against the $G_5^{(2)}$ terms as we will discuss at the end of the next section.

\section{Coefficient of $\ln(s/\mu^2)$ }

\begin{figure}[h]
    \includegraphics{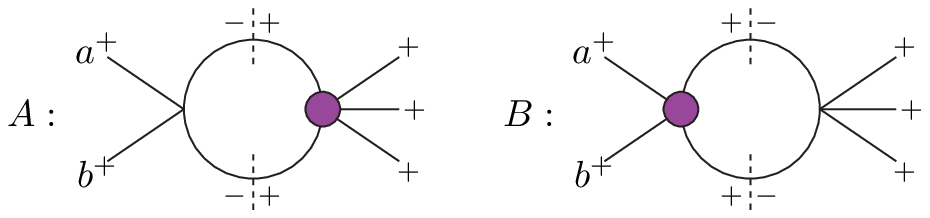}
  \caption{Contributions to the two-loop amplitudes involving an all-plus loop (indicated by $\bullet$)}
    \label{fig:fivepointbubbles}
\end{figure}

We determine the presence of the $\ln(s_{ab}/\mu^2)$ functions using two-particle cuts.  The coefficient has two contributions as shown
in fig.\ref{fig:fivepointbubbles}.

We determine these using canonical forms.
The canonical basis approach~\cite{Dunbar:2009ax} is a systematic method to determine the coefficients of triangle and bubble integral functions in a one-loop
amplitude from
the three and two-particle cuts.  
\noindent{A} two particle cut is of the form
\begin{equation}
C_{2}  \equiv  i   \int d^4 \ell_1  \delta( \ell_1^2)\delta(\ell_2^2) 
A^{\rm tree}_1(-\ell_1, a, \cdots, b, \ell_2) \times A^{\rm tree}_2(-\ell_2,\cdots, \ell_1)\,.
\end{equation}
The product of tree amplitudes appearing in the two-particle cut can be decomposed 
in terms of  canonical forms ${\cal H}_i$, 
\begin{equation}
A^{\rm tree}_1(-\ell_1, \cdots, \ell_2) \times A^{\rm tree}_2(-\ell_2,\cdots, \ell_1)  = 
\sum e_i {\cal H}_i (\rho_k, {\ell_j}),
\end{equation} 
where the $e_i$ are coefficients independent of $\ell_j$.  
We then use
substitution rules to replace the ${\cal H}_i (\rho_k, {\ell_j})$ by 
rational functions $H_i[\rho_k,P ]$  to obtain the coefficient of the bubble integral function as
\begin{equation}
\sum_i e_i H_i[ \rho_k , P] \,.
\end{equation}
Here we use this technique treating the one-loop all-plus vertex as a tree amplitude.

The canonical forms we need and their substitutions are
\def\BB{B}
\def\BA{A}
\def\BC{C}
\def\TB{b}
\def\TA{a}
\def\TC{c}
\def\K{P}
\begin{align}
{\cal H}^0_{1,1} (A,C ;  a,c ,  \ell_1,\ell_2  )  &  \equiv   
{\la \TA \ell_1 \ra \over \la \BA \ell_1 \ra}
{\la \TC \ell_2 \ra \over \la \BC \ell_2\ra }
\longrightarrow
{H}^0_{1,1} [A,C ;  a,c; P ]= 
{\spa{C}.c [C|P|a\ra \over \spa{C}.{A} [C|P|C\ra }
+
{\spa{A}.c [A|P|a\ra \over \spa{A}.{C} [A|P|A\ra }  
\notag 
\\
{\cal H}_{2x}^0 ( A; B_1 ; C_1; \ell_1,\ell_2  )& \equiv{\spa{B_1}.{\ell_1}\spa{C_1}.{\ell_2}\over \spa{A}.{\ell_1}\spa{A}.{\ell_2} }
\longrightarrow
{ H}_{2x}^0 [ A; B_1 ; C_1 ; P ]= { [A|P|B_1\ra [A|P|C_1\ra\over [A|P|A\ra^2 }.
\label{eq:canonicalforms} 
\end{align}
where $P$ is the cut-momentum $P=k_a+\cdots k_b$. 
We take the product of amplitudes in the cut and split them into a sum of terms of type in eqn.(\ref{eq:canonicalforms}) using
partial fractioning
\begin{equation}
{ \prod_{j=1}^{n-1}  \spa{\ell}.{X_j} \over \prod_{i=1}^n \spa{\ell}.{Y_i}  }=
\sum_{i=1}^{n} \left( \prod_{j=1}^{n-1} \spa{Y_i}.{X_j} \over \prod_{l\neq i} \spa{Y_l}.{Y_{i} } \right)    \times { 1 \over \spa{\ell}.{Y_i} }
= \sum_{i=1}^{n} \alpha_i { 1\over \spa{\ell}.{Y_i} } 
\end{equation}

For the five-point amplitude the cuts have $P=k_a+k_b$.  Working specifically with  $s_{ab}=s_{45}$, 
the two-particle cut has two contributions
\begin{align}
A:& M_5^{(0)}(1^+,2^+,3^+,l_2^-,l_1^-) \times M_4^{(1)}(4^+,5^+,l_1^+,l_2^+) 
\notag 
\\
B:& M_5^{(1)}(1^+,2^+,3^+,l_2^+,l_1^+) \times M_4^{(0)}(4^+,5^+,l_1^-,l_2^-) 
\end{align}
From the term A we obtain a contribution to the coefficient of
\begin{align}
C^A_{45}=\sum_{r=1,2} c_r \sum_{i=1}^3\sum_{j=1,j\neq i}^2  \alpha^r_{i}\beta^r_j   H^0_{1,1}[ A_i^r,a_j^r; B_3^r,b_3^r; P]
+\sum_{r=1,2} c_r \sum_{i=1}^2   \alpha^r_{i}\beta^r_i   H^0_{2x}[ A_i^r; B_3^r,b_3^r; P]
\end{align}
with
\begin{align}
\alpha^r_i = { \prod_{l=1}^2 \spa{B_l^r}.{A_i^r}  \over \prod_{l\neq i} \spa{A_l^r}.{A_i^r} }
\;\;\;
\beta^r_i = {  \spa{b_1^r}.{a_i^r}  \over \prod_{l\neq i} \spa{a_l^r}.{a_i^r} }
\end{align}
where
\begin{equation}
c_1 =-{ \spb4.5^6 \spb2.3 \over 120 \spa2.3\spa1.3 } 
\;\;\;
c_2={ \spb4.5^6 \spb1.2 \over 120 \spa1.2\spa1.3 } 
\end{equation}
and
\begin{align}
&\{ |A_i^1\ra \}  =\{   |1\ra , |2\ra , |3\ra \} ,\{ |A_i^2\ra \} =\{   |2\ra , |3\ra , |1\ra \} 
\notag 
\\
&\{ |a_i^1\ra \} =\{   |1\ra , |2\ra  \} \; , \; \{ a_i^2 \}=\{   |2\ra , |3\ra \} 
\notag 
\\
&\{ |B_i^1\ra \} = \{ |5\ra , |5\ra , -K_{45}|1] \}  \;\; \{ |B_i^2\ra \} = \{ |5\ra , |5\ra , -K_{45}|3] \} 
\notag 
\\
&\{ |b_i^1\ra \} = \{ |b_i^2 \} = \{ |4\ra , |4\ra  \} 
\end{align}

Contributions from the second configuration are more complicated, but using  relationships between the different terms to simplify the final expression
we obtain 
\begin{equation}
C^B_{45}= T^B_{1,2,3,4,5}+T^B_{2,1,3,4,5}+T^B_{3,2,1,4,5}  
\end{equation}
with
\begin{align}
T^B_{1,2,3,4,5}= { \spb2.3  \spb4.5
 \over 120  \spa2.3 \spa4.5
 } \left(  
\sum_{i=2}^5 \sum_{j=2,j \neq i }^5 \alpha_i \beta_j  H_{1,1}^0[i,j; A_4,B_4,\{4,5\}]
+\sum_{i=2}^5 \alpha_i \beta_i  H_{2x}^0[i ; A_4,B_4,\{4,5\}]
\right)
\end{align}
where
\begin{align}
\alpha_i = {\prod_{k=1}^3 \spa{i}.{A_k} \over \prod_{l\in \{2,3,4,5\}-\{i\}} \spa{i}.{l} } 
\;\;\;
\beta_j = {\prod_{k=1}^3 \spa{j}.{B_k} \over \prod_{l\in \{2,3,4,5\}-\{j\} } \spa{j}.{l} } 
\end{align}
and
\begin{align}
\{ |A_k\ra \} = \{ |B_k\ra \} =\{ |1\ra  , K|1] , K|1] , K|1] \} 
\end{align}

Thus the amplitude contains
\begin{equation}
(C^A_{45}+C^B_{45}) \left( {1 \over \epsilon} -\ln( s_{45}/\mu^2 ) \right)
\end{equation}
The full amplitude thus has $\epsilon^{-1}$ and $\ln(\mu^2)$ terms of
\begin{equation}
\sum_{i<j} ( C^A_{ij}+C_{ij}^B )    \ln(s_{ij}) +
\sum_{i<j} ( C^A_{ij}+C_{ij}^B ) \left( \frac{1}{\epsilon}+\ln(\mu^2 ) \right)
\label{eq:fivepointinfinity}
\end{equation}
We have determined the above expression using four-dimensional unitarity which isolates the coefficient of $\ln(s_{ij})$. The value of 
$\ln(\mu^2)$ then follows.  The coefficient of $\epsilon^{-1}$ in this is tied to the $\ln(\mu^2)$ but is, presumably, not the value which would be obtained from 
a field theory calculation.

The bubble coefficients contain spurious $\spa{e}.c^{-3}$ singularities which must cancel~\cite{Dunbar:2011xw}  against the singularities in the $F_5^{(2)}$ term. 
Specifically the singularities are of the form
\begin{equation}
{ 1 \over \spa{e}.c^{3} }   \times \left\{  \ln( s_{ab} ),  \ln( s_{dc}) , \ln (s_{ed})  +\hbox{\rm permutations} \right\}  
\end{equation}
where the permutations are of of $a,b,d$. 
The $F_5^{(2)}$ term contains dilogarithms but near the point $\spa{e}.c=0$ these simplify and
\begin{equation}
F_5^{(2)} = { 1\over \spa{e}.c^4 } \biggl( 0 + s_{ec} \left\{  \ln( s_{ab} ),  \ln( s_{dc}) , \ln (s_{ed}) +\hbox{\rm permutations} \right\} 
+{\cal O} (s_{ec}^2 ) \biggr) 
\end{equation}
We have explicitly checked that within $F_5^{(2)}+G_5^{(2)}$ both the $\spa{e}.c^{-3}$ {\it and } $\spa{e}.c^{-2}$ singularities cancel, leaving the full amplitude free of spurious singularities: this is a strong consistency check. We have also checked the collinear limit of the five-point amplitude. 

\section{Counterterm Lagrangian}

In this section we enumerate the possible independent counterterms for pure gravity in
four dimensions.
In general, graviton scattering amplitudes,
in $D$ dimensions at $L$ loops,
require the introduction
of counterterms of the
form
\begin{equation}
\nabla^n R^m
\end{equation}
where $n+2m = (D-2)L+2$
and
we have suppressed the indices on $R$.
$R$ may stand either for
the Riemann tensor, $R_{abcd}$, the Ricci tensor $R_{ab}\equiv
g^{cd}R_{acbd}$ or the curvature scalar $R \equiv g^{ab}R_{ab}$.
Although, there are  a large number of tensor structures which may
appear, fortunately, the symmetries of the Riemann tensor reduce these
considerably. Firstly, there are the basic symmetries of $R_{abcd}$
\begin{equation}
R_{abcd}=-R_{bacd}=-R_{abdc}=R_{cdab}
\end{equation}
and the cyclic symmetry,
\begin{equation}
R_{abcd}+R_{acdb}+R_{adbc}=0
\end{equation}
Secondly, we have the Bianchi identity for $\nabla_e R_{abcd}$,
\begin{equation}
\nabla_e R_{abcd}
+\nabla_c R_{abde}
+\nabla_d R_{abec}
=0
\end{equation}
There are also ``derivative identities'' which involve two covariant derivatives,
\begin{align}
\nabla_{e}\nabla_{f} R_{abcd}
-\nabla_{f}\nabla_{e} R_{abcd}
&= R^g{}_{aef} \, R_{gbcd}
+ R^g{}_{bef} \, R_{agcd}
+ R^g{}_{cef} \, R_{abgd}
+ R^g{}_{def} \, R_{abcg}
\cr
\nabla^2 R_{abcd}
& =
 2 \, R^{f}_{\ a c e} \, R^{e}_{\ d b f}
- 2 \, R^{f}_{\ b c e} \, R^{e}_{\ d a f}
- R^{e}_{\ d a b} \, R_{c e}
+ R^{e}_{\ c a b} \, R_{d e}
\cr
& \hskip 1cm
+ \nabla_{c} \nabla_{a} R_{b d} - \nabla_{c} \nabla_{b} R_{a d}
- \nabla_{d} \nabla_{a} R_{b c} + \nabla_{d} \nabla_{b} R_{a c}
\label{DerivativeSymmetry}
\end{align}

\noindent
These symmetries will be used to determine the minimal set of
inequivalent counterterms.

From power counting the possible two-loop
 counterterms in $D=4$ are of the form $R^3$ or $\nabla^2 R^2$.
The independent terms involving $R_{abcd}$, $R_{ab}$ and $R$
are \cite{R3D4,Fulling:1992vm},
\begin{align}
T_1&=\nabla_{a} R \nabla^{a} R
&   T_2=&\nabla_{a} R_{bc} \nabla^{a} R^{bc}
\cr
T_3&=\nabla_{e} R_{abcd} \nabla^{e} R^{abcd}
& T_4=&\nabla_{c} R_{ab} \nabla^{b} R^{ac}
\cr
T_5&= R^3
& T_6=& R R_{ab}  R^{ab}
\cr
T_7&= R R_{abcd}  R^{abcd}
& T_8= &R_{abcd} R^{abce} R^{d}_{\ e}
\cr
T_9&=  R_{abcd} R^{ac} R^{bd}   
& T_{10}= & R_{a}{}^b R_b{}^cR_c{}^a
\cr
T_{11} &= R^{ab}{}_{cd}R^{cd}{}_{ef}R^{ef}{}_{ab}  \;\;\;
& T_{12}= & R_{abcd}R^{a}{}_e{}^c{}_f R^{bedf}
\end{align}
For the case of pure gravity, the counterterm structure can be
represented as a single counterterm with a numerical coefficient. We
review the argument leading to the conclusion that a single
counterterm is sufficient.
(When matter is
coupled to gravity this is no longer the case.)

For pure gravity the equation of motion is
\begin{equation}
R_{ab}=0
\end{equation}
Hence terms involving the Ricci tensor or curvature scalar
will not contribute to the $S$-matrix and such terms can be
discarded when calculating the counterterms. (If calculating an off-shell
object, such counterterms can, and do, appear.)
Ignoring such terms leaves us with three tensors - $T_3$, $T_{11}$ and $T_{12}$.
The term $T_3$
\begin{equation}
T_3=\nabla_e R_{abcd}
\nabla^e R^{abcd} \equiv-  R_{abcd} \nabla^2 R^{abcd}
\end{equation}
can be rearranged using the  identity
in eq.~(\ref{DerivativeSymmetry})
into terms involving the Ricci tensor plus
cubic terms in the Riemann tensor.
Thus for pure gravity this term is equivalent to a combination of
$T_{11}$ and $T_{12}$ and can be
eliminated from the list of inequivalent counterterms.

Finally, in six dimensions the scalar topological density can be written
\begin{equation}
\delta^{\ a \, b \, c d e f}_{[mnpqrs]} R^{mn}{}_{ab} R^{pq}{}_{cd}R^{rs}{}_{ef}
\end{equation}
which implies that the combination
\begin{equation}
\sum_{i=5}^{12} a_i T_i \equiv 0
\end{equation}
is topological for some coefficients $a_i$ in dimensions $D\leq 6$.
Hence for pure gravity amplitudes we can replace
$T_{12}$ with $T_{11}$ (or vice versa).  Thus we are led to the fact that
the counterterm can be taken as a single tensor with a coefficient.
Thus the counterterm can be chosen to be
\begin{equation}
{ C_{R^3} \over 60} \times ({\kappa \over 2})^2{1 \over (4\pi)^4 } \int d^4 x \sqrt{-g} \; R^{ab}{}_{cd} R^{cd}_{ef} R^{ef}{}_{ab}
\end{equation}  
with the free coefficient $C_{R^3}$.

Computing with this Lagrangian, the parts of the four-point amplitudes proportional to $C_{R^3}$ 
are~\cite{Dunbar:2002jx}
\begin{align}
{\cal M}^{R^3}_4(1^+,2^+,3^+,4^+) &=  C_{R^3} \times \left( {\kappa \over 2}\right)^6 \times { 1 \over (4\pi)^4}
\left( { s t \over \spa1.2\spa2.3\spa3.4\spa4.1 } \right)^2 { stu }
\notag
\\
{\cal M}^{R^3}_4(1^-,2^+,3^+,4^+) &= C_{R^3} \times \left( {\kappa \over 2}\right)^6 \times { 1 \over (4\pi)^4}  
\left( { \spb2.4^2 s^2  t \over \spb1.2\spa2.3\spa3.4\spb4.1 } \right)^2 
{ t \over 10 s u }
\notag
\\
{\cal M}^{R^3}_4(1^-,2^-,3^+,4^+) &= C_{R^3} \times 0
\end{align}
Comparing this to the coefficient of $\ln(\mu^2)$ in eqn.~(\ref{eq:fourpoint}) we find 
\begin{equation}
C_{R^3}= -\frac{1}{4}
\end{equation}

We also require the five point amplitude computed using the above Lagrangian. Perturbative gravity calculations based upon Feynman diagrams 
are notoriously difficult, however we can compute the higher point functions using recursion. 

The original BCFW shift~\cite{Britto:2005fq}, 
\begin{equation}
\lambda_i \longrightarrow \lambda_i+z\lambda_j \;\;, \;
\bar\lambda_j \longrightarrow \bar\lambda_j -z\bar\lambda_i \;\;
\label{eq:BCFshift}
\end{equation}
does not lead to an expression with the correct symmetry, however the shift~\cite{Risager:2005vk,BjerrumBohr:2005jr}
\begin{align}
\lambda_a\to &\lambda_{\hat a} = \lambda_a +z\spb{b}.c \lambda_\eta \,,
\notag \\
\lambda_b\to &\lambda_{\hat b} = \lambda_b +z\spb{c}.a \lambda_\eta \,,
\notag \\
\lambda_c\to &\lambda_{\hat c} = \lambda_c +z\spb{a}.b \lambda_\eta \,,
\label{KasperShift}
\end{align}
where $\lambda_\eta$ is an arbitrary spinor does.  Using this shift we can obtain an expression for the five-point amplitude.  The amplitude has 
two factorisations,
\begin{align}
M_3^{\tree}( \hat a^+, d^+, K^-)  &\times {1\over K^2} \times M_4^{R^3}( \hat b^+ , \hat c^+ , d^+ , -K^+) \;\;\; 
\notag \\
M_3^{\tree}( \hat b^+, \hat c^+, K^-)  &\times {1\over K^2} \times M_4^{R^3}(  \hat a^+ , d^+, e^+ , -K^+)
\end{align}
with six terms of the first type and three of the second. With $(a,b,c,d,e)=(1,2,3,4,5)$ we find that the first term gives
\begin{align}
T^A_{1,2,3,4,5}=N_4 { \spb1.4 \over \spa1.4 } 
{ \spb5.3  \spb5.2 \over \spa1.{\eta}^2\spa4.{\eta}  }{ \spb2.3^2
\over  \spa4.5 }\times 
  [5|K_{14}|\eta\ra 
 [2|K_{14}|\eta \ra 
 [3|K_{14}|\eta \ra  
\end{align}
which is symmetric under $2\leftrightarrow 3$. 
The second is
\begin{align}
T^B_{1,2,3,4,5}=-N_4  
{ \spb1.4 \spb1.5 \spb2.3 [1|K_{23}|\eta\ra^2 [5|K_{23}|\eta\ra  [4|K_{23}|\eta\ra \over \spa2.3 \spa2.\eta^2\spa3.\eta^2  }
 {  \spb4.5 \over  \spa4.5   }
\end{align}
which is symmetric under $2\leftrightarrow 3$ and $4\leftrightarrow 5$.
The normalisation is
\begin{equation}
N_4= \left( {\kappa \over 2}\right)^6 \times { 1 \over (4\pi)^4} \times C_{R^3} 
\end{equation}
The resultant contribution to the amplitude is
\begin{equation}
{\cal M}^{R^3}_5(1^+,2^+,3^+,4^+,5^+) = N_4 \left( \sum_{P_6}  T^A_{1,2,3,4,5}+\sum_{P_3} T^B_{1,2,3,4,5} \right)
\label{eq:rcubedfivepoint}\end{equation}
where the summation is over the six independent $T^A$ and the three independent $T^B$. 

\noindent
The expression for ${\cal M}^{R^3}_5(1^+,2^+,3^+,4^+,5^+)$ is

$\bullet$ 
Fully crossing symmetric between external legs

$\bullet$
Independent of the spinor $\eta$

\noindent
These are strong indicators that we have computed the correct expression. We have checked this construction for the $n$-point all-plus amplitude up to $n=8$. 

\noindent
Additionally

$\bullet$  
As $z \longrightarrow \infty$ for the BCFW shift the amplitude does not vanish but behaves as $z^{2}$.  This is the reason why the shift (\ref{eq:BCFshift}) does not generate this amplitude. 

$\bullet$  The expression has soft limits with
\begin{equation}
{\cal M}_5^{R^3} 
=(  {1\over t^3} S^{(0)} +{1\over t^2} S^{(1)} +{1\over t
} S^{(2)} ) {\cal M}^{R^3}_{4} +O(t^0)
\end{equation} 
where $S^{(i)}$ are the leading, sub-leading and sub-sub-leading soft operators~\cite{SoftTheorems}.  As a two-loop amplitude, there is a possibility that the sub-sub-leading would not satisfy this so we regard this as a feature of the constructed amplitudes rather than as a necessary constraint. 

Comparing expression (\ref{eq:rcubedfivepoint}) with (\ref{eq:fivepointinfinity}) 
we find
\begin{equation}
\left( \sum_{P_6}  T^A_{1,2,3,4,5}+\sum_{P_3} T^B_{1,2,3,4,5} \right) =-4\times \sum_{i<j} ( C^A_{ij}+C_{ij}^B )
\end{equation}
and
we therefore obtain $C_{R^3}=-1/4$. The counterterm is thus consistent with that required 
for the four point amplitude.

\section{Conclusion}

Computing quantum gravity amplitudes is notoriously difficult. Only a small number of on-shell scattering amplitudes have been computed analytically.
For pure gravity only the four and five-point one-loop 
amplitudes have been presented for all helicity configurations with all-$n$ expressions for the all-plus and single-minus amplitudes. Progress beyond one-loop has been confined to theories which are supersymmetric where the enhanced symmetries significantly simplify the amplitudes.

In this article we have shown how four dimensional cutting techniques allow us to compute large and interesting parts of two-loop pure gravity amplitudes and have
obtained the (poly)logarithmic parts of the all-plus helicity amplitude for four and five-points in compact analytic forms.  
We also obtain the associated $\ln(\mu^2)$ terms which as argued in ref.~\cite{Bern:2015xsa} determine the non-renormalisability of the amplitudes. 
We have matched these to 
the same $R^3$ counterterm for both the four and five-point amplitude. Given that the $\ln(\mu^2)$ terms are key to renormalisability, this technique provides a straightforward method to study the UV behaviour of gravity theories. 

Our approach has been entirely based upon physical on-shell amplitudes and is very different from a field theory approach where the one-loop renormalisation uses ``evanescent'' operators~\cite{Bern:2015xsa}. We do not obtain the $\epsilon^{-1}$ term found there, but do reproduce the four-point {\it renormalised}
amplitude and present a five-point amplitude correctly renormalised. 

{\it Note added} As this article was been prepared ref.~\cite{Bern:2017puu} appeared where the four-point two-loop amplitude is studied using similar techniques. 

\section{Acknowledgements}

This work was supported by Science and Technology Facilities Council (STFC) grant ST/L000369/1.


\begin{thebibliography}{99}

\bibitem{Eden}
R.J. Eden, P.V. Landshoff, D.I. Olive, J.C. Polkinghorne, {\it
The Analytic S Matrix}, (Cambridge University Press, 1966).

\bibitem{Badger:2013gxa}
  S.~Badger, H.~Frellesvig and Y.~Zhang,
  JHEP {\bf 1312} (2013) 045
  doi:10.1007/JHEP12(2013)045
  [arXiv:1310.1051 [hep-ph]].

\bibitem{Gehrmann:2015bfy} 
  T.~Gehrmann, J.~M.~Henn and N.~A.~Lo Presti,
  Phys.\ Rev.\ Lett.\  {\bf 116}, no. 6, 062001 (2016)
  Erratum: [Phys.\ Rev.\ Lett.\  {\bf 116}, no. 18, 189903 (2016)]
  doi:10.1103/PhysRevLett.116.189903, 10.1103/PhysRevLett.116.062001
  [arXiv:1511.05409 [hep-ph]].





\bibitem{Dunbar:2016aux}
  D.~C.~Dunbar and W.~B.~Perkins,
  Phys.\ Rev.\ D {\bf 93} (2016) no.8,  085029
  doi:10.1103/PhysRevD.93.085029
  [arXiv:1603.07514 [hep-th]].



\bibitem{Bern:1994zx}
  Z.~Bern, L.~J.~Dixon, D.~C.~Dunbar and D.~A.~Kosower,
  Nucl.\ Phys.\ B {\bf 425} (1994) 217
  [hep-ph/9403226].
  
 
\bibitem{Bern:1994cg}
  Z.~Bern, L.~J.~Dixon, D.~C.~Dunbar, D.~A.~Kosower,
  Nucl.\ Phys.\  {\bf B435 } (1995)  59
  [hep-ph/9409265].
  
\bibitem{Dunbar:2016cxp}
  D.~C.~Dunbar, G.~R.~Jehu and W.~B.~Perkins,
  Phys.\ Rev.\ D {\bf 93} (2016) no.12,  125006
  doi:10.1103/PhysRevD.93.125006
  [arXiv:1604.06631 [hep-th]].

 
\bibitem{Dunbar:2016gjb}
  D.~C.~Dunbar and W.~B.~Perkins,
  Phys.\ Rev.\ Lett.\  {\bf 117} (2016) no.6,  061602
  doi:10.1103/PhysRevLett.117.061602
  [arXiv:1605.06351 [hep-th]].


\bibitem{Badger:2016ozq}
  S.~Badger, G.~Mogull and T.~Peraro,
  JHEP {\bf 1608} (2016) 063
  doi:10.1007/JHEP08(2016)063
  [arXiv:1606.02244 [hep-ph]].

\bibitem{OneLoopFinite}
G. 't Hooft and M. Veltman,
Ann.\ Poincare\ Phys.\ Theor.\ {\bf A20}:69 (1974) \\
G. 't Hooft, Nucl.\ Phys.\ {\bf B62}:444 (1973) 



\bibitem{Goroff:1985th}
  M.~H.~Goroff and A.~Sagnotti,
  Nucl.\ Phys.\ B {\bf 266} (1986) 709.
  doi:10.1016/0550-3213(86)90193-8


\bibitem{vandeVen:1991gw}
  A.~E.~M.~van de Ven,
  Nucl.\ Phys.\ B {\bf 378} (1992) 309.
  doi:10.1016/0550-3213(92)90011-Y


\bibitem{Bern:2015xsa}
  Z.~Bern, C.~Cheung, H.~H.~Chi, S.~Davies, L.~Dixon and J.~Nohle,
  Phys.\ Rev.\ Lett.\  {\bf 115} (2015) no.21,  211301
  doi:10.1103/PhysRevLett.115.211301
  [arXiv:1507.06118 [hep-th]].

\bibitem{Weinberg:1965nx}
  S.~Weinberg,
  Phys.\ Rev.\  {\bf 140} (1965) B516.
  doi:10.1103/PhysRev.140.B516
  
  

\bibitem{Dunbar:1995ed}
  D.~C.~Dunbar and P.~S.~Norridge,
  Class.\ Quant.\ Grav.\  {\bf 14} (1997) 351
  doi:10.1088/0264-9381/14/2/009
  [hep-th/9512084].
  
\bibitem{Naculich:2011ry}
  S.~G.~Naculich and H.~J.~Schnitzer,
  JHEP {\bf 1105} (2011) 087
  doi:10.1007/JHEP05(2011)087
  [arXiv:1101.1524 [hep-th]].

\bibitem{Akhoury:2011kq}
  R.~Akhoury, R.~Saotome and G.~Sterman,
  Phys.\ Rev.\ D {\bf 84} (2011) 104040
  doi:10.1103/PhysRevD.84.104040
  [arXiv:1109.0270 [hep-th]].


\bibitem{Witten:2003nn}
  E.~Witten,
  Commun.\ Math.\ Phys.\  {\bf 252} (2004) 189
  [hep-th/0312171].


\bibitem{Catani:1998bh}
  S.~Catani,
  Phys.\ Lett.\ B {\bf 427} (1998) 161
  doi:10.1016/S0370-2693(98)00332-3
  [hep-ph/9802439].
   

\bibitem{SWI}
M.T.\ Grisaru, H.N.\ Pendleton and P.\ van Nieuwenhuizen,
Phys.\ Rev.\ {\bf D15}:996 (1977) \\
M.T. Grisaru and H.N. Pendleton, Nucl.\ Phys.\ {\bf B124}:81 (1977) \\
S.J. Parke and T. Taylor, Phys.\ Lett.\ {\bf B157}:81 (1985)

 
 
\bibitem{Dunbar:1994bn}
  D.~C.~Dunbar and P.~S.~Norridge,
  Nucl.\ Phys.\ B {\bf 433} (1995) 181
  doi:10.1016/0550-3213(94)00385-R
  [hep-th/9408014].



\bibitem{Bern:1998sv}
  Z.~Bern, L.~J.~Dixon, M.~Perelstein and J.~S.~Rozowsky,
  Nucl.\ Phys.\ B {\bf 546} (1999) 423
  doi:10.1016/S0550-3213(99)00029-2
  [hep-th/9811140].

\bibitem{Berends:1988zp}
  F.~A.~Berends, W.~T.~Giele and H.~Kuijf,
  Phys.\ Lett.\ B {\bf 211} (1988) 91.
  doi:10.1016/0370-2693(88)90813-1

\bibitem{Bern:1996ja}
  Z.~Bern, L.~J.~Dixon, D.~C.~Dunbar and D.~A.~Kosower,
  Phys.\ Lett.\ B {\bf 394} (1997) 105
  doi:10.1016/S0370-2693(96)01676-0
  [hep-th/9611127].


\bibitem{Bern:1995db}
  Z.~Bern and A.~G.~Morgan,
  Nucl.\ Phys.\ B {\bf 467} (1996) 479
  doi:10.1016/0550-3213(96)00078-8
  [hep-ph/9511336].


\bibitem{BrittoUnitarity} R.~Britto, F.~Cachazo and B.~Feng,
  Nucl.\ Phys.\ B {\bf 725} (2005) 275 [hep-th/0412103].



\bibitem{Bern:1993kr}
  Z.~Bern, L.~J.~Dixon and D.~A.~Kosower,
  Nucl.\ Phys.\ B {\bf 412} (1994) 751
  doi:10.1016/0550-3213(94)90398-0
  [hep-ph/9306240].


\bibitem{Bidder:2005ri}
  S.~J.~Bidder, N.~E.~J.~Bjerrum-Bohr, D.~C.~Dunbar and W.~B.~Perkins,
  Phys.\ Lett.\  B {\bf 612} (2005) 75
  [hep-th/0502028].



\bibitem{Darren}
  D.~Forde,
  Phys.\ Rev.\ D {\bf 75} (2007) 125019
  [arXiv:0704.1835 [hep-ph]].


\bibitem{BjerrumBohr:2007vu}
  N.~E.~J.~Bjerrum-Bohr, D.~C.~Dunbar and W.~B.~Perkins,
  JHEP {\bf 0804} (2008) 038
  [arXiv:0704.1835 [hep-ph]].



\bibitem{Mastrolia:2006ki}
  P.~Mastrolia,
  Phys.\ Lett.\  B {\bf 644} (2007) 272
  [arXiv:hep-th/0611091].
  

  
\bibitem{Dunbar:2009ax}
  D.~C.~Dunbar, W.~B.~Perkins and E.~Warrick,
  JHEP {\bf 0906} (2009) 056
  [arXiv:0903.1751 [hep-ph]].
  

\bibitem{Dunbar:2011xw}
  D.~C.~Dunbar, J.~H.~Ettle and W.~B.~Perkins,
  Phys.\ Rev.\ D {\bf 84} (2011) 125029
  doi:10.1103/PhysRevD.84.125029
  [arXiv:1109.4827 [hep-th]].


\bibitem{R3D4}
R.E.~Kallosh,
Nucl. Phys. {\bf B78}:293 (1974) \\
P.~van Nieuwenhuizen and C.C.~Wu,
J. Math. Phys. {\bf 18}:182 (1977)


\bibitem{Fulling:1992vm}
  S.~A.~Fulling, R.~C.~King, B.~G.~Wybourne and C.~J.~Cummins,
  Class.\ Quant.\ Grav.\  {\bf 9} (1992) 1151.
  doi:10.1088/0264-9381/9/5/003


\bibitem{Dunbar:2002jx}
  D.~C.~Dunbar and N.~W.~P.~Turner,
  Phys.\ Lett.\ B {\bf 547} (2002) 278
  doi:10.1016/S0370-2693(02)02759-4
  [hep-th/0203104].



\bibitem{Britto:2005fq}
  R.~Britto, F.~Cachazo, B.~Feng and E.~Witten,
  Phys.\ Rev.\ Lett.\  {\bf 94} (2005) 181602
  [hep-th/0501052].


\bibitem{Risager:2005vk}
  K.~Risager,
  JHEP {\bf 0512} (2005) 003
  doi:10.1088/1126-6708/2005/12/003
  [hep-th/0508206].

\bibitem{BjerrumBohr:2005jr}
  N.~E.~J.~Bjerrum-Bohr, D.~C.~Dunbar, H.~Ita, W.~B.~Perkins and K.~Risager,
  JHEP {\bf 0601} (2006) 009
  doi:10.1088/1126-6708/2006/01/009
  [hep-th/0509016].


\bibitem{SoftTheorems}
Steven Weinberg, 
Phys.Rev. 140 (1965) B516-B524;\\
  C.~D.~White,
  JHEP {\bf 1105} (2011) 060
  [arXiv:1103.2981 [hep-th]];\\
F.~Cachazo and A.~Strominger,
arXiv:1404.4091 [hep-th];\\
  Z.~Bern, S.~Davies and J.~Nohle,
  Phys.\ Rev.\ D {\bf 90}, no. 8, 085015 (2014)
  doi:10.1103/PhysRevD.90.085015
  [arXiv:1405.1015 [hep-th]];\\
  S.~D.~Alston, D.~C.~Dunbar and W.~B.~Perkins,
  Phys.\ Rev.\ D {\bf 92}, no. 6, 065024 (2015)
  doi:10.1103/PhysRevD.92.065024
  [arXiv:1507.08882 [hep-th]].

\bibitem{Bern:2017puu}
  Z.~Bern, H.~H.~Chi, L.~Dixon and A.~Edison,
  arXiv:1701.02422 [hep-th].


\end{thebibliography}
\end{document}